\documentclass[pre,twocolumn,showpacs,amsmath,amssymb]{revtex4}
\usepackage{graphicx} 

\newcommand{\grad}{\boldsymbol{\nabla}}
\newcommand{\Hext}{\ensuremath{\textbf{H}_{\text{ext}}}}
\renewcommand{\eqref}[1]{Eq.~(\ref{#1})}

\begin{document}

\title{Microfluidic capturing-dynamics of paramagnetic bead suspensions}

\author{Christian Mikkelsen and Henrik Bruus}

\affiliation{MIC -- Department of Micro and Nanotechnology,\\
Technical University of Denmark, DK-2800 Kongens Lyngby, Denmark}

\date{6 May 2005}
\begin{abstract}

We study theoretically the capturing of paramagnetic beads by a
magnetic field gradient in a microfluidic channel treating the
beads as a continuum. Bead motion is affected by both fluidic and
magnetic forces. The transfer of momentum from beads to the fluid
creates an effective bead-bead interaction that greatly aids
capturing. We demonstrate that for a given inlet flow speed a
critical density of beads exists above which complete capturing
takes place.
\end{abstract}

 \maketitle
\section{Introduction}
\label{sec:introduction} Recently, there has been an increasing
interest in using magnetic beads in separation of, say,
biochemical species in microfluidic
systems \cite{Choi:00a,Pankhurst:03a}. The principle is to have
biochemically functionalized polymer beads with inclusions of
superparamagnetic nanometersize particles of, for example,
magnetite or maghemite. They attach to particular biochemical
species and can be separated out from solution by applying
external magnetic fields. As most biological material is either
diamagnetic or weakly paramagnetic, this separation is
specific. Paramagnetic particles in fluids are also used to
measure the susceptibility of, for example, magnetically labelled
cells by measuring particle capture or motion in a known
field \cite{Zborowski:95a,McCloskey:03a}.

In this paper we study microfluidic capturing of paramagnetic
beads from suspension by modeling the beads as a continuous
distribution \cite{Warnke:03a}. The separation of suspended
paramagnetic beads from their host fluid is an important process
as it decides operating characteristics for practical microfluidic
devices. It involves an interplay between forces of several kinds
governing the dynamics of the process: (a) Magnetic forces from
the application of strong magnetic fields and field gradients. (b)
Drag forces due to the motion of the beads with respect to the
host fluid. (c) The trivial effect of gravity, which we ignore in
the following. We emphasize the effects of bead motion on the
fluid flow as this gives rise to a hydrodynamic interaction
between the beads. As we have noted in a previous few-bead study,
this interaction is more important than the magnetic bead-bead
interactions \cite{Mikkelsen:05a}. It is created by drag forces in
two steps: First, drag transfers momentum to the fluid from the
beads moving under the influence of external forces. Second, the
modified flow changes the drag on and thus motion of other beads.

\begin{figure}[b]
\centerline{\includegraphics[]{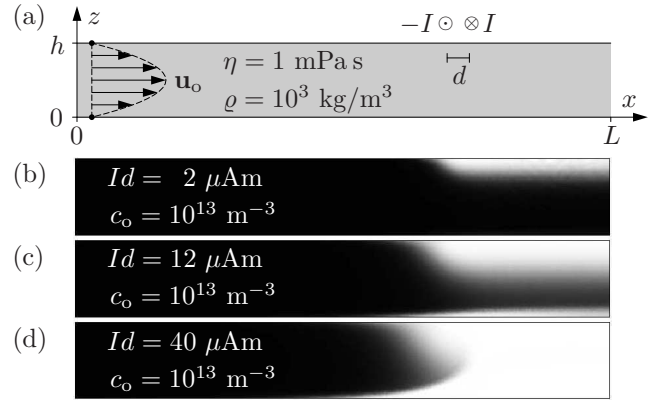}} \caption{(a)
\label{fig:GeometryConc} Sketch of the microfluidic system with
$L=350~\mu$m and $h=50~\mu$m. A suspension of paramagnetic beads
enters at $x=0$ with a parabolic Poiseuille flow profile, $\mathbf{u}_{\text{o}}$,
and leaves at $x=L$. Beads are caught by the
pair of wires placed $100~\mu$m from the outlet at the top and
carrying currents $\pm I$. (b)--(d) Simulated stationary density
of the beads ranging from zero (white) to $c_\text{o}$ (black) for
increasing values of the current-distance product $Id$ as
indicated. At $x=0$ the concentration  is $c_\text{o} =
10^{13}$~m$^{-3}$ and the maximum flow speed is $300~\mu$m/s.}
\end{figure}

\section{Model}
As sketched in Fig.~\ref{fig:GeometryConc}(a) we consider a
viscous fluid (water) flowing in the $x$ direction between a pair
of parallel, infinite, planar walls. The walls are placed parallel
to the $xy$ plane at $z=0$ and $z=h$, respectively. A steep
magnetic field gradient is generated by a parallel pair of closely
spaced, infinitely long, thin wires along the $y$ direction
separated by $d$ and carrying opposite currents $\pm I$. The
system is translation invariant in the $y$ direction thereby
reducing the simulation to a tractable problem in 2D. The
simulation domain is defined by $0<x<L$ and $0<z<h$ with $L =
350~\mu$m and $h = 50~\mu$m. The wires intersect the $xz$ plane
near $(x,z)=(250~\mu\text{m},\ 55~\mu\text{m})$ just above the top
plate. Paramagnetic beads in suspension are injected into the
microfluidic channel by the fluid flow at $x=0$. They are either
exiting the channel at $x=L$ or getting collected at the channel
wall near the wires.

When a suspension of beads is viewed on a sufficiently large scale
compared to the single bead radius $a$ but on a scale comparable
to density variations, we can describe the distribution of beads
in terms of a continuous, spatially varying bead number density
$c$. We consider a suspension of beads with radius $a=1~\mu$m and
denote the initial number density at $x=0$ by $c_{\text{o}}$. The
four basic constituents of the model are described in the
following.

\emph{Magnetic force.} The beads are paramagnetic with a magnetic
susceptibility $\chi=1$. In an external magnetic field
$\Hext(\textbf{r})$ the force on such a bead is
 \begin{align}
 \textbf{F}_{\text{ext}} & = \mu_\text{o}\!\int_{\text{bead}}
 (\textbf{M}\cdot\grad)\Hext \,dV\nonumber \\
 \label{eq:magforce}
 & = 4\pi\mu_\text{o} a^3 \frac{\chi}{\chi+3}(\Hext\cdot\grad)\Hext
 \end{align}
assuming that the bead is so small that we can take the external
field $\Hext$ to be approximately constant across the bead, i.e.,
$a |\grad\Hext| \ll |\Hext|$ when determining the magnetization
$\textbf{M}$.

As mentioned, $\Hext$ in this study arises from a pair of current
carrying wires. It is determined in the following manner. From
Amp\`{e}re's law, we readily find the magnetic field,
$\textbf{H}$, around a straight circular wire, $\textbf{H}
(\textbf{r}) = \textbf{J}\times\textbf{r}/(2\pi r^2)$, where the
electrical current vector $\textbf{J}$ is along the wire
orthogonal to the position vector $\textbf{r}$ which is in the
$xz$ plane. The magnetic field from the two closely spaced
anti-parallel wires is found by decreasing the separation $d$ and
increasing the current, $I=|\textbf{J}|$, while keeping the
product $I d$ constant,
 \begin{equation} \textbf{H}_{\text{ext}}
 = \frac{1}{2\pi r^2}\left(\textbf{J}\times\textbf{d} -
 \frac{2(\textbf{J}\times\textbf{r})(\textbf{d}\cdot\textbf{r})}{r^2}\right).
 \end{equation}
This together with \eqref{eq:magforce} yields
 \begin{equation}  \textbf{F}_{\text{ext}} = -\frac{2}{\pi}
 \frac{\chi}{\chi+3}\mu_\text{o}
 a^3 (I\, d)^2 \,\frac{\textbf{r}}{r^6}\: ;
 \end{equation}
a manifestly attractive central force (from the mid-point of the
wires), independent of the direction of $\textbf{d}$.

\emph{Fluid flow.} \label{Sec:flow} The beads are suspended in a
fluid of viscosity $\eta$ and density $\varrho$ that is launched
at $x=0$ with a parabolic velocity profile, $\mathbf{u}_{\text{o}}$, and
flows past the wires. In microfluidics inertial effects are
unimportant compared to drag, so the small beads in suspension
almost always move with constant velocity relative to the fluid.
Except for acceleration phases shorter than microseconds the
external forces are exactly balanced by drag \cite{Newton2}. The
momentum transfer from beads to fluid is included by adding a bulk
force term, $c\textbf{F}_{\text{ext}}(\textbf{r})$ to the
Navier--Stokes equation. This bulk force term is proportional to
the number density $c$ of beads and the magnetic external force
$\textbf{F}_{\text{ext}}$ on an individual bead at position
$\textbf{r}$. The velocity $\textbf{u}$ of the fluid is given by
 \begin{equation}  \varrho\partial^{{}}_t \textbf{u}+\varrho(\textbf{u}\cdot\grad){\textbf{u}}
 = -\grad p +\eta\nabla^2 \textbf{u} + c\textbf{F}_{\text{ext}},
 \label{eq:NS}
 \end{equation}
along with the incompressibility condition $\grad\cdot\textbf{u} =
0$.

\emph{Bead motion.} \label{Sec:beads} To complete the set of
equations, it is necessary to have an equation of motion for the
bead number density $c$. As the beads neither appear nor disappear
in the bulk, $c$ must obey a continuity equation
 \begin{equation}  \partial^{{}}_t c + \grad\cdot\textbf{j} = 0,
 \label{eq:particlecontinuity}
 \end{equation}
where the bead current \textbf{j} is defined by the Nernst--Planck
equation \cite{Probstein:94}
 \begin{equation} \label{eq:NP}
 \textbf{j} = -D\grad c + c
 \textbf{u} + cb\textbf{F}_{ \text{ext}}
 \end{equation}
with diffusivity $D$ and bead mobility $b = 1/(6\pi\eta a)$.

For our spherical beads the diffusivity is given by the Einstein
expression $D = kT/(6\pi\eta a)$ which for water at room
temperature equals $2.2\times 10^{-13}~$m$^2$/s. In the
simulations below, however, we artificially increase the magnitude
of $D$ in order to stabilize the computations and to use a coarser
mesh and thus save computation time.

\begin{figure}[t]
\centerline{\includegraphics[width=\columnwidth]{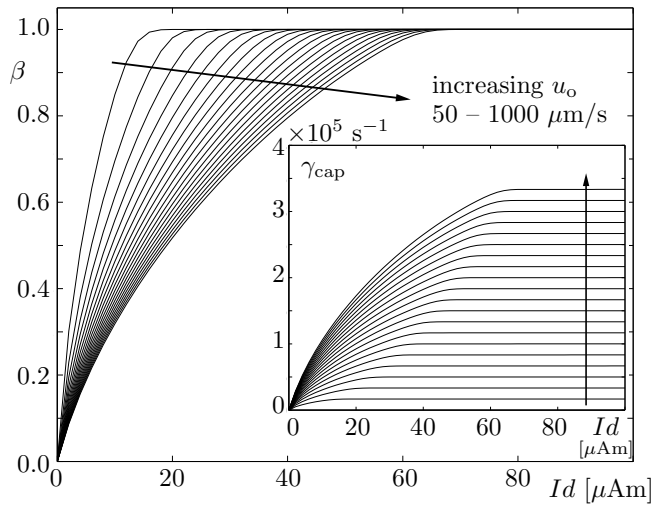}}
\caption{\label{fig:BetaGamma} The fraction $\beta$ of beads
caught as function of the current-distance product $Id$ for twenty
different flow speeds (50 -- 1000~$\mu$m/s; indicated by the
arrows). Larger current leads to higher $\beta$; faster flow to
smaller $\beta$. In this simulation the initial concentration is
low, $c_\text{o} = 10^{13}$~m$^{-3}$. \emph{Inset:} Rate
$\gamma_{\text{cap}}$ of bead capture as function of $Id$, for the
flows above. The faster flow or the larger current, the higher
$\gamma_{\text{cap}}$.}
\end{figure}

\emph{Boundary conditions.} In addition to the bulk
equations~(\ref{eq:NS}), (\ref{eq:particlecontinuity}),
and~(\ref{eq:NP}), we need appropriate boundary conditions. As the
beads move out to the walls of the domain and settle there, merely
demanding that the normal component of the bead current vector
$\textbf{j}$ vanishes is not correct, rather, it must be free to
take on any value as long as it is directed into the wall. As
beads do not enter the bulk from the walls (by assumption once
settled, beads stick) we demand that the normal current component
is never directed into the liquid. For the fluid we demand the
usual no-slip condition at the walls.

At the inlet $x=0$ of the microfluidic channel we assume that the
fluid comes in with the constant initial number density
$c_{\text{o}}$ and with a parabolic fluid velocity profile
$\mathbf{u}_{\text{o}}$. At the outlet $x=L$ we let the bead
current take on any value, while the fluid pressure is zero.

\section{Results}
\label{Sec:results}\setcounter{paragraph}{0}

Having set up the equations for bead and fluid motion, they are
solved with the finite element method on a mesh with $\sim 10^4$
elements refined in the vicinity of the wires. To this end we
employ the finite element solver software package
Femlab \cite{Femlab}. The parameter values for the fluid are those
of water, $\eta = 1$~mPa$\:$s and $\varrho = 10^3$~kg/m$^3$, while
for the beads $a=1~\mu$m and $c_{\text{o}} = 10^{13}$ to
$10^{16}$~m$^{-3}$.

To study capturing we must keep track of which beads are captured
and which are flushed through the channel with the flow. This is
done by calculating the rates $\gamma_i$ by which the beads are
either captured or transported in/out at each of the four boundary
segments $i$ of the channel (inlet, outlet, upper wall, and lower
wall). By integration of the normal components of the bead
currents along each segment $i$, we find
 \begin{equation}
 \gamma_i=\int_i\textbf{j}\!\cdot\!\textbf{n}\: d\ell_i.
 \end{equation}
The rate of capture is $\gamma_{\text{cap}} =
\gamma_{\text{lower}} + \gamma_{\text{upper}}$. In steady state
the conservation of beads enforces $\gamma_{\text{inlet}} +
\gamma_{\text{cap}} + \gamma_{\text{outlet}} = 0$, which provides
a useful check of the simulation results. The primary control
parameters are the current-wire distance product $Id$, the maximum
fluid in-flow speed $u_{\text{o}}$, and the bead number density,
$c_{\text{o}}$. The product $Id$ decides the magnetic force which
captures the beads against the fluid flow. As we are investigating
effects of bead-bead interaction, our interest is properties that
depend on the bead number density, in particular those that do so
nonlinearly.

\begin{figure}[t]
\centerline{\includegraphics[width=\columnwidth]{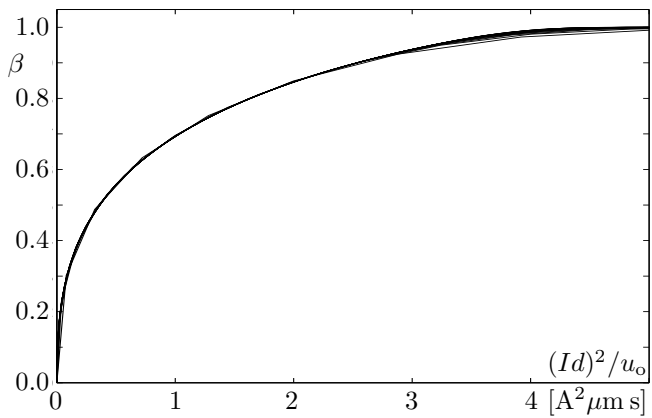}}
\caption{\label{fig:collapsed} The fraction $\beta$ of beads
caught versus $(Id)^2/u_{\text{o}}$, the ratio of the
current-distance product squared and the fluid flow velocity. This
demonstrates scaling in the competition between capturing and
flushing: the twenty curves from Fig.~\ref{fig:BetaGamma}
approximately collapse to one single.}
\end{figure}

\emph{Electrical current and fluid flow.} The effects of having
electrical wires near, and thus a magnetic field gradient in, the
channel is illustrated in Fig.~\ref{fig:GeometryConc}(b)--(d) for
three values of the current-distance product $Id$. At small values
of $Id$ only a narrow region is emptied but increasing the current
the region expands until it covers the width of the channel.

A simple measure of the capturing is the ratio $\beta$ of the bead
capture rate $\gamma_{\text{cap}}$ to the bead in-flow rate
$\gamma_{\text{inlet}}$,
 \begin{equation} \beta =
 \frac{\text{``capture rate''}}{\text{``in-flow rate''}} =
 \frac{\gamma_{\text{cap}}}{\gamma_{\text{inlet}}}.
 \end{equation}
If capturing dominates $\beta$ tends to unity, if flushing
dominates $\beta$ tends to zero. Fig.~\ref{fig:BetaGamma} shows
this in that slow flow and strong current leads to a high $\beta$
whereas fast flow and weak current leads to a small value. The
rate $\gamma_{\text{cap}}$ of bead capture as function of wire
current and flow velocity is illustrated in the inset of
Fig.~\ref{fig:BetaGamma}.

If there is a competition between magnetic capturing and flushing,
then we expect that the data can be described essentially by the
ratio of the magnetic forces to the inlet fluid flow speed
$u_{\text{o}}$. The force is proportional to the square of the
current-distance product $Id$. In Fig.~\ref{fig:collapsed}, we
plot the data from Fig.~\ref{fig:BetaGamma} as function of
$(Id)^2/u_{\text{o}}$ and see that the data mostly collapses onto
a single curve. The collapse is not perfect and is not expected to
be as the underlying flow and bead distribution patterns (see
Fig.~\ref{fig:GeometryConc}) are different for different flows and
magnetic fields.

\begin{figure}[t]
\centerline{\includegraphics[width=\columnwidth]{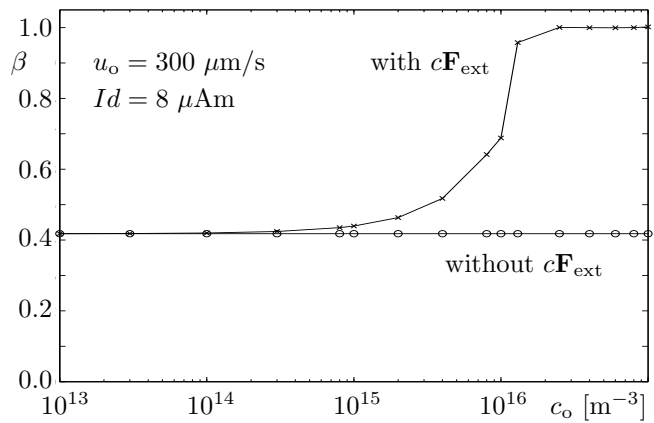}}
\caption{\label{fig:captureconc} Fraction $\beta$ of beads caught
as function of initial bead number density with and without the bulk force
term $c\textbf{F}_\text{ext}$ in \eqref{eq:NS}. The fixed values
for the current-distance product $Id$ and the maximum in-flow
speed $u_{\text{o}}$ are shown. At low densities less than 50\%\
are caught; at high densities the collective motion of the beads
leads to 100\% capture.}
\end{figure}

\emph{Interactions and concentration.} The second point we wish to
make is that modification of the overall flow, and the effective
bead-bead interaction this entails, is significant for bead
capturing. We can study the effect by excluding momentum transfer
to the fluid flow due to the bulk force term
$c\textbf{F}_\text{ext}$ in the Navier--Stokes
equation~(\ref{eq:NS}). At high bead number densities the force
acting on the beads contributes a significant force on the fluid
affecting fluid flow and spawning the effective interaction. The
strength of this interaction must thus depend on the density of
particles. This is illustrated in Fig.~\ref{fig:captureconc};
capturing was simulated at fixed in-flow speed,
$u_{\text{o}}=300~\mu$m/s, and a fixed value of the current-wire
distance product, $Id=8~\mu$Am, but for varying bead number
densities $c_ \text{o}$ ranging from $10^{13}$ to
$10^{16}$~m$^{-3}$. At low densities we find that capturing is
roughly independent of density and the fraction $\beta$ of beads
captured has some intermediate value, however, for high densities
all beads are caught. In contrast, leaving out the bulk force term
$c\textbf{F}_{\text{ext}}$ in the Navier--Stokes equation, i.e.,
the force acting on the fluid, gives concentration independence as
shown in Fig.~\ref{fig:captureconc}.

\begin{figure}[t]
\centerline{\includegraphics[width=\columnwidth]{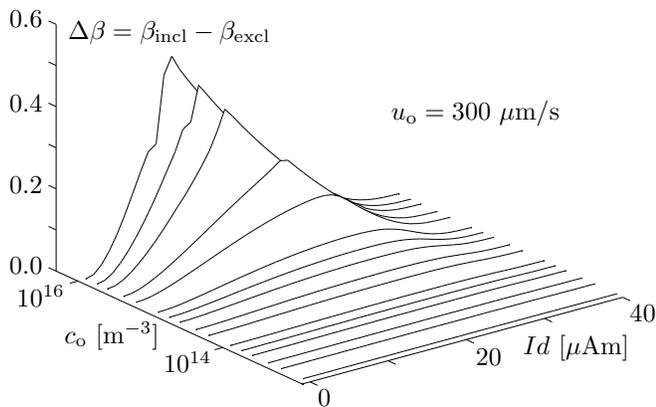}}
\caption{\label{fig:DeltaBeta} The difference $\Delta\beta =
\beta_{\text{incl}} - \beta_{\text{excl}}$ in captured bead
fractions between two situations: including and excluding the bulk
force term $c\textbf{F}_{\text{ext}}$ in the Navier--Stokes
equation~(\ref{eq:NS}). At high concentrations ($c_\text{o}
> 10^{15}$~m$^{-3}$) there is an appreciable difference between
including and excluding the bulk force term, corresponding to
hydrodynamic bead-bead interactions. }
\end{figure}

As can be seen in Fig.~\ref{fig:DeltaBeta}, a complementary way of
exhibiting the importance of the bulk force term is to plot the
difference $\Delta \beta = \beta_\text{incl} - \beta_\text{excl}$
between including and excluding $c\textbf{F}_{\text{ext}}$ as a
function of concentration and the current-distance product. This
shows that interactions makes an appreciable difference at high
concentrations and intermediate magnetic fields.

\emph{Diffusion constant.} Even for the small beads of radius
$a=1~\mu$m, the diffusion constant given by the Einstein relation
is small compared to the dimensions entering the problem. The
time-scale for a bead to diffuse across the channel is
$\tau_{\text{diff}}\sim h^2/D$. If we are to see the influence of
diffusion competing with bead advection, then the relevant
quantity is the P\'{e}clet number $h u/D$ which is advection
time-scale $\tau_{\text{adv}}\sim h/u$ over the diffusion
time-scale. When this number is larger than unity, which it is
except for artificially large diffusion constants, then convection
dominates. In the simulations the diffusion constant is increased
artificially up to $10^{-11}$~m$^2$/s in order to help numerical
convergence. But we have verified that values smaller than
$10^{-10}$~m$^2$/s have no influence on the results.

\section{Discussion and conclusion}
We have studied microfluidic capture of paramagnetic beads in
suspension. The three main findings of work are: the approximate
scaling shown in Fig.~\ref{fig:collapsed}, the existence of a
critical bead density for capture shown in
Fig.~\ref{fig:captureconc}, and the qualitative difference for
capturing between models with and without the hydrodynamic
bead-bead interaction shown in Figs.~\ref{fig:captureconc}
and~\ref{fig:DeltaBeta}.

Clearly, it is very important for the capture process to include
the action of the beads on the host fluid medium. Simply leaving
it out can give qualitatively wrong results for high
concentrations of beads. This casts some doubt on the measurement
of cell susceptibility through capturing as it depends on cell
concentration \cite{Zborowski:95a,McCloskey:03a}. Deduction of
susceptibilities from single bead or cell considerations together
with measurements at high bead or cell concentration is suspect.
Care must be taken to compare with standards of known and similar
susceptibility, size, and concentration.

The effective bead-bead interaction  greatly helps capturing. It
should make detectable differences depending on whether there are
a few or hundreds of particles in a channel at a time in actual
experiments especially when the flow and magnetic field are such
that the beads are barely caught one by one. This interaction
should be considered when choosing operating conditions for
microfluidic devices based on capturing of beads as higher bead
number densities potentially eases requirements for external
magnets and allows faster flushing. We hope that experimental
studies will be initiated to verify this prediction of our work.

\textbf{Acknowledgements}. We thank Mikkel Fougt Hansen and
Kristian Smistrup for valuable discussions on magnetophoresis in
general and of their experiments in particular.


\end{document}